\documentclass{vgtc}                          

\ifpdf
  \pdfoutput=1\relax                   
  \usepackage{graphicx}                
  \DeclareGraphicsExtensions{.pdf,.png,.jpg,.jpeg} 
\else
  \usepackage{graphicx}                
  \DeclareGraphicsExtensions{.eps}     
\fi%

\graphicspath{{figures/}{pictures/}{images/}{./}} 

\usepackage{microtype}                 
\PassOptionsToPackage{warn}{textcomp}  
\usepackage{textcomp}                  
\usepackage{mathptmx}                  
\usepackage{times}                     
\usepackage{cite}                      
\usepackage{tabu}                      
\usepackage{booktabs}                  
\usepackage{subcaption}

\usepackage[frozencache=true,cachedir=minted-cache]{minted} 
\setminted{
    fontsize=\scriptsize,
    baselinestretch=1,
    breaklines=0 . 
}

\usepackage{xspace,xpunctuate}
\newcommand{\ie}{{i.e.,}\xspace}
\newcommand{\eg}{{e.g.,}\xspace}
\newcommand{\ea}{{et~al\xperiod}\xspace}

\newcommand{\etc}{{etc\xperiod}\xspace}

\usepackage{enumitem}
\usepackage{hyperref}

\onlineid{1018}

\vgtccategory{System or Tool}

\title{Draco~2: An Extensible Platform to Model Visualization Design}

\author{
Junran Yang\thanks{e-mail: junran@cs.washington.edu}\\ %
    \scriptsize University of Washington,\\ \scriptsize Seattle %
\and Péter Ferenc Gyarmati\thanks{e-mail: peter.ferenc.gyarmati@univie.ac.at}\\ %
    \scriptsize University of Vienna,\\ \scriptsize Vienna %
\and Zehua Zeng\thanks{e-mail: zhzeng@umd.edu} \\ %
    \scriptsize University of Maryland,\\ \scriptsize College Park %
\and Dominik Moritz\thanks{e-mail: domoritz@cmu.edu}\\ %
     \scriptsize Carnegie Mellon University,\\ \scriptsize Pittsburgh %
}

\abstract{
Draco introduced a constraint-based framework to model visualization design in an extensible and testable form.
It provides a way to abstract design guidelines from theoretical and empirical studies and applies the knowledge in automated design tools.
However, Draco is challenging to use because there is limited tooling and documentation.
In response, we present Draco~2, the successor with (1) a more flexible visualization specification format, (2) a comprehensive test suite and documentation, and (3) flexible and convenient APIs.
We designed Draco~2 to be more extensible and easier to integrate into visualization systems.
We demonstrate these advantages and believe that they make Draco~2 a platform for future research.
} 

\CCScatlist{
  \CCScatTwelve{Human-centered computing}{Visualization}{Visualization systems and tools}{}
}

\begin{document}


\firstsection{Introduction}
\maketitle

For the remainder of this paper, we refer to the original system as Draco~1, while Draco and Draco~2 denote our presented contribution.

Draco~1~\cite{Moritz2018formalizing} aims to make design guidelines concrete, actionable, and testable.
By encoding guidelines as logical rules, developers and researchers can build computational knowledge bases for automatically assessing existing charts~\cite{hopkins2020visualint,9552878,McNutt2018LintingFV} and generating new recommended charts~\cite{Wongsuphasawat2015voyager,Wongsuphasawat2017voyager2,Wongsuphasawat2016towards,Mackinlay1986automating}.
To evaluate and recommend charts, Draco~1 uses a constraint solver that checks whether a specification is valid or finds a design that incurs a minimal cost of constraint violations.
Using this approach, Draco~1 can warn users about ineffective designs and provide design suggestions, even for partial and ambiguous requests.
By formalizing design guidelines as constraints, visualization researchers and practitioners can add their own design considerations and immediately test the results, seeing how different rules and weights change which visualizations are considered the most preferable.
The weights of the constraints in the Draco~1 knowledge base can be adjusted by hand or learned from graphical perception results \cite{Zeng2023review,Zeng2021evaluation,Zhu2020survey} using machine-learning~\cite{Hu2019vizml,Li2022kg4vis}.



Since its introduction in 2018, Draco~1 has become a popular tool in the visualization research community. Yet, even though Draco~1 as a conceptual framework can keep up with expanding design rules over time, its original implementation and knowledge base are limited in two major ways.
First, Draco~1 only focuses on single views and faceted views (using row and column encodings) limiting its support for multi-view visualizations. Its specification language is also closely tied to Vega-Lite~\cite{Satyanarayan2017vegalite}. It is not generalizable to render recommended results with other visualization libraries. 
Second, Draco~1 comes with limited documentation, low test coverage, and limited tooling support, making it difficult to adopt it as a library. 
The inability to be integrated into emerging projects hinders Draco~1 from being a true platform for future research in visualization recommendation and reasoning about visualization design.

To overcome these limitations, we introduce Draco, an improved system for capturing and applying visualization design best practices. In summary:

\begin{itemize}[nosep,leftmargin=*]
    \item Draco introduces an \textit{improved visualization specification format}, independent of Vega-Lite, which makes it more flexible and extensible. It supports much more designs, including multi-layer and multi-view visualizations and makes scales a first-class concept. 
    \item Draco (1) has \textit{thorough documentation} covering the core modules, lower-level building blocks, and examples, and (2) a comprehensive test suite with \textit{100\% unit test coverage}, making it more reliable and suitable for adoption. Draco is easier to set up as it runs entirely in Python (the original Draco~1 system needed Python and JavaScript). We provide a \textit{REST API} so Draco can still be integrated into web-based applications. We also distribute Draco as a \textit{WebAssembly} package which runs in any modern browser.
    \item Draco has convenient APIs to interact with the knowledge base to convert constraints from and to a nested format, validate specifications, recommend optimal completions of partial specifications (visualization recommendation), and to debug, adapt, and extend the knowledge base. We consider the default knowledge base in Draco as a starting point that researchers and systems builders adapt using these tools.
\end{itemize}


We believe that our work is a significant update to the original Draco~1 system. We envision that with the improvements described in this paper, Draco~2 can be a solid platform for customization and future visualization research. Towards this goal, we make Draco~2 available as open source at \href{https://github.com/cmudig/draco2}{github.com/cmudig/draco2}.
In this paper, we discuss the version \href{https://github.com/cmudig/draco2/releases/tag/v2.0.0}{2.0.0}.




\section{Background and Related Work}



Draco upgrades its previous version, Draco~1~\cite{Moritz2018formalizing}---a framework for efficiently modeling visualization design knowledge towards constructing new visualization recommendation algorithms, with a more general visualization specification and more user-friendly utilities for knowledge representation. 


\textbf{Visualization Specification:}
Automated visualization tools use specification languages to describe and synthesize visualization designs. Mackinlay’s APT~\cite{Mackinlay1986automating} system ranks encoding choices based on the \textit{expressiveness} and \textit{effectiveness} criteria. Tableau's ShowMe~\cite{Mackinlay2007showme} suggests specific encodings with heuristic rules. Voyager’s~\cite{Wongsuphasawat2015voyager,Wongsuphasawat2017voyager2} CompassQL~\cite{Wongsuphasawat2016towards} and Draco~1 both build on the Vega-Lite~\cite{Satyanarayan2017vegalite} grammar and combine rules that model fine-grained design knowledge with hand-tuned scores. In Draco, we re-design the logical representation to a generalized and extended chart specification format that is extensible and renderer-agnostic. With such a format, we support multiple views and view composition. 

\textbf{Modeling Visualization Design Knowledge:} 
Visualization recommendation researches on algorithms including rule-based methods considering theoretical principles~\cite{Mackinlay1986automating,Mackinlay2007showme,Wongsuphasawat2015voyager,Wongsuphasawat2017voyager2} or proposing new metrics~\cite{Vartak2015seedb,Demiralp2017foresight,Key2012vizdeck}, and ML-based approaches~\cite{Hu2019vizml,Li2022kg4vis,Luo2018deepeye} learning from a vast corpus of empirical results. 
Visualization recommendation frameworks~\cite{Wongsuphasawat2016towards,Siddiqui2017fast,Moritz2018formalizing} have been proposed to make it easier to design and test new recommendation algorithms. Draco~1 is the only one that was designed with the explicit goal of being extensible and adaptable. It allows modeling visualization design knowledge in Answer Set Programming (ASP) and uses the Clingo solver~\cite{Gebser2011potassco,Gebser2011asp,Gebser2014clingo} to search the constrained space and rank the answers with weighted costs. 
Draco aims to provide a platform for future research beyond constructing visualization recommendation algorithms. For example, Zeng \ea~\cite{Zeng2023toomanycooks} use Draco to analyze the implication of different graphical perception studies.


\section{Components of Draco}

Draco has three main components: a general description language for charts, a knowledge base that encodes best practices using hard and soft constraints, and an API to manipulate the knowledge base and programmatically reason about the knowledge base using a constraint solver.

\subsection{Draco Specification Format}


To express knowledge over visualization designs, Draco describes visualizations as logical \textit{facts} similar to Draco~1.
Draco specifies charts in a more generic and extensible way than its previous version. 
In general, it describes the structure of charts as nested specifications. Since Clingo needs a flat list of facts, Draco has methods to convert the nested representation to (\mintinline{python}{dict_to_facts}) and from (\mintinline{python}{answer_set_to_dict}) a flat list of logical facts with the relationships also expressed as logical facts.



\subsubsection{Nested specification format with Entities and Attributes}


\begin{figure}[htbp]
    \centering
    \includegraphics[width=\columnwidth]{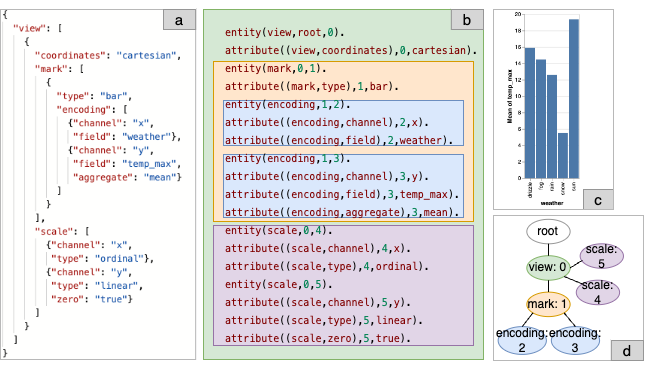}
    \caption{(a) Chart specification as a nested dictionary (left, here as a Python \mintinline{python}{dict}), (b) flattened list of logical facts (middle, in Answer Set Programming), (c) the rendered visualization result, and (d) the skeleton abstracted from the logical format.}
    \label{fig:spec_example}
\end{figure}

Draco specifies the nested chart specification with two kinds of facts: \textit{entity} and \textit{attribute} as shown in \autoref{fig:spec_example}. Entities describe objects and their association with unique identifiers, while attributes describe the properties of entities. For example, \mintinline{prolog}{entity(mark,v0,m1). attribute((mark,type),m1,bar).} specifies a mark object \texttt{m1} of type \texttt{bar} on the view object \texttt{v0}. 

The logical format can be seen as a tree where the entities and attributes are the nodes connected by the entity keys. For example, \autoref{fig:spec_example} (d) shows the skeleton of the specification with only the entities. 
Using this format, Draco allows for both complete and partial specification as input. A complete specification has attributes associated with each entity, specifying a ready-to-render chart. A partial specification defines either part of all components in a complete chart (\eg one layer of a multi-layer chart), the skeleton of a chart without the attributes (\eg a single-view-single-layer chart with an arbitrary mark type), or a mix of both. Our enumeration algorithm supported by ASP augments the skeleton with additional entities and completes it by filling in potential attributes. Meanwhile, Draco reasons about fine-grained rules related to subtrees of the input by checking if they satisfy the given constraints. Additional hints can be included as part of the query to constrain the target outputs. For example, adding aggregate rules like \mintinline{prolog}{:- {entity(encoding,_,_)} <= 2.} filters out designs with less than three encodings (headless rules are integrity constraints that derive false from their body, and satisfying the body results in a contradiction, which is disallowed). The tree structure of logical format allows for querying, searching, and reasoning about abstract visualization composition.

The dictionary format is an abstraction from the logical format that can generalize to multiple visualization specification languages with customized renderers. Its compact format keeps the structural information but is agnostic of entity identifiers. Thus, it deduplicates structural equivalent specifications whose entity keys are different. For instance, \mintinline{prolog}{entity(view,root,v0). attribute((view,coords,v0,polar))} and \mintinline{prolog}{entity(view,root,0). attribute((view,coords,0,polar))} would both be represented as \mintinline{json}{{"view":[{"coords":"polar"}]}} in a dictionary format since they are structurally identical, even though the entity keys \mintinline{latex}{v0} and \mintinline{latex}{0} are different.

\subsubsection{Encoding Visualizations in Draco}

Draco~1 uses an encoding based on the Grammar of Graphics (GoG)~\cite{wilkinson2005gog} and Vega-Lite~\cite{Satyanarayan2017vegalite}. However, Draco~2 is not limited to the features Vega-Lite supports. 
For instance, in Vega-Lite and Draco~1, each encoding has a data type that describes the semantics of the data (quantitative, temporal, ordinal, or nominal). However, encoding data types are omitted in Draco~2 because they can be automatically inferred from the combination of primitive field type (number, string, etc.) and the scale type (linear, log, ordinal, or categorical) for the encoding.
Therefore, Draco~2's constraints directly reason about primitive field type and scale type, which are both explicit elements in visualizations compared to encoding type.
Draco~2 also makes scales an explicit entity that can be associated and shared with multiple encodings across marks.
Shared scales allow for comparisons across marks. Vega-Lite, which does not explicitly make scales independent entities, has to resort to a mechanism in which authors specify how scales in a view resolve\footnote{\href{https://vega.github.io/vega-lite/docs/resolve.html}{vega.github.io/vega-lite/docs/resolve.html}}.

Because of the nested format with entities and attributes, Draco~2 is generic and convenient to extend.  
\autoref{fig:spec_example} shows the complete specification for a single-view single-layer bar chart. However, a Draco~2 program can encode visualizations in multiple views, where a view can contain one or more marks that encode data and corresponding scales. 
If a view has multiple marks, Draco~2 assumes that the marks are in the same view space in the chart (\ie layered).
Besides the visualization, a Draco~2 program can describe the data schema and the primary visualization task.
This format could easily be extended with additional attributes and entities. For instance, one could add additional attributes about a mark such as the font size, scale, or color scheme. New entities could be legends and axes so that Draco~2's constraints could then reason about the position, size, or other properties of these guides.

\subsection{Knowledge Base}
Just like its predecessor, Draco uses a collection of hard and soft constraints over the logical facts to represent design knowledge guidelines. While the hard constraints span the space of all designs considered valid, the soft constraints define the preferences over the space to rank these designs. 
When a user queries with a partially specified visualization, Draco eliminates ill-formed (\eg using the encoding channel \textit{shape} for a mark that is not \textit{point}) or non-expressive (\eg aggregating \textit{ordinal} fields with \textit{mean} function) designs with the hard constraints and searches the design space for the lowest-cost specifications. The \textit{Draco cost} of a visualization is the weighted sum of the costs of all violated soft constraints. And the weights reflect the relative importance of each violation in the total cost. As a starting point, the default knowledge base consists of constraints adopted from CompassQL rules. 

There are two ways to obtain soft constraint weights.
First, our API allows algorithm designers to define their own sets of soft constraints and manually assign a weight to each constraint to indicate their preferences.
Second, Draco-Learn can learn weights for existing soft constraints from ranked pairs of visualizations (the learning algorithm is the same as in Draco~1~\cite{Moritz2018formalizing}). These pairs could come from different experimental studies or theoretical rankings.

Draco loads and exposes the knowledge base as answer-set programs. We use the \textbf{definitions} programs to declare the domains of visualization attributes. For example, \mintinline{prolog}{domain((mark,type),} \mintinline{prolog}{(point;bar;line;area;text;tick;rect)).} defines the choices of mark types. To enforce the search space to follow the correct Draco general description language, we use the \textbf{constraints} programs. \mintinline{prolog}{violation(invalid_domain) :- } \mintinline{prolog}{attribute(P,_,V), domain(P,_), not domain(P,V). }, for example, allows only valid domain values following the definitions. Then, we have a generator from the \textbf{generate} programs that sets up the search space. For example, the rule
\vspace{-0.5em}
\begin{minted}[breaklines]{prolog}
{ attribute((N,A),E,V): domain((N,A),V) } = 1 :- entity(N,_,E), required((N,A)).
required((mark,type)).
\end{minted}
\vspace{-0.5em}
requires every mark to have one type from its domain. Finally, Draco loads the hard and soft constraints as \textbf{hard} and \textbf{soft} programs. 
Default soft constraint weights are declared in a separate file, and can be loaded and assigned to the constraints when a \mintinline{python}{Draco}  object is instantiated, which also allows for customized weights.
Each program is a dictionary of \emph{blocks} consisting of the constraint and its description. Blocks allow users to pick and choose parts of a program to filter the knowledge base and access documentation. We include unit tests for every constraint and the parser that reads the constraints and documentation from ASP into a Python dictionary.

\subsection{Other API and Development Tooling}


We provide a well-documented API for interacting with Draco's knowledge base, encapsulating core utilities for use, extension, and customization.
To maintain code quality and to make Draco a viable platform for future research, we follow engineering best practices (\ie static type analysis, code linting, enforcement of 100\% unit test coverage, \etc). Here, we present our Python API and briefly discuss the development tooling we provide. Comprehensive documentation can be found at \href{https://dig.cmu.edu/draco2}{dig.cmu.edu/draco2}.


\textbf{Specification Renderer:} 
The Draco specification format provides an abstract and machine-readable way to express visualizations. Although, rendered visualization are necessary to efficiently communicate the specifications. 
We present a default Vega-Lite-based~\cite{Satyanarayan2017vegalite} renderer.
However, the Draco specification format is generalized and not limited to specific visualization grammars. 
Therefore, our renderer is extensible such that it comes with an interface from which custom rendering logic can be implemented with the help of our supporting documentation. 

\begin{figure}[t]
    \centering
    \includegraphics[width=0.475\textwidth]{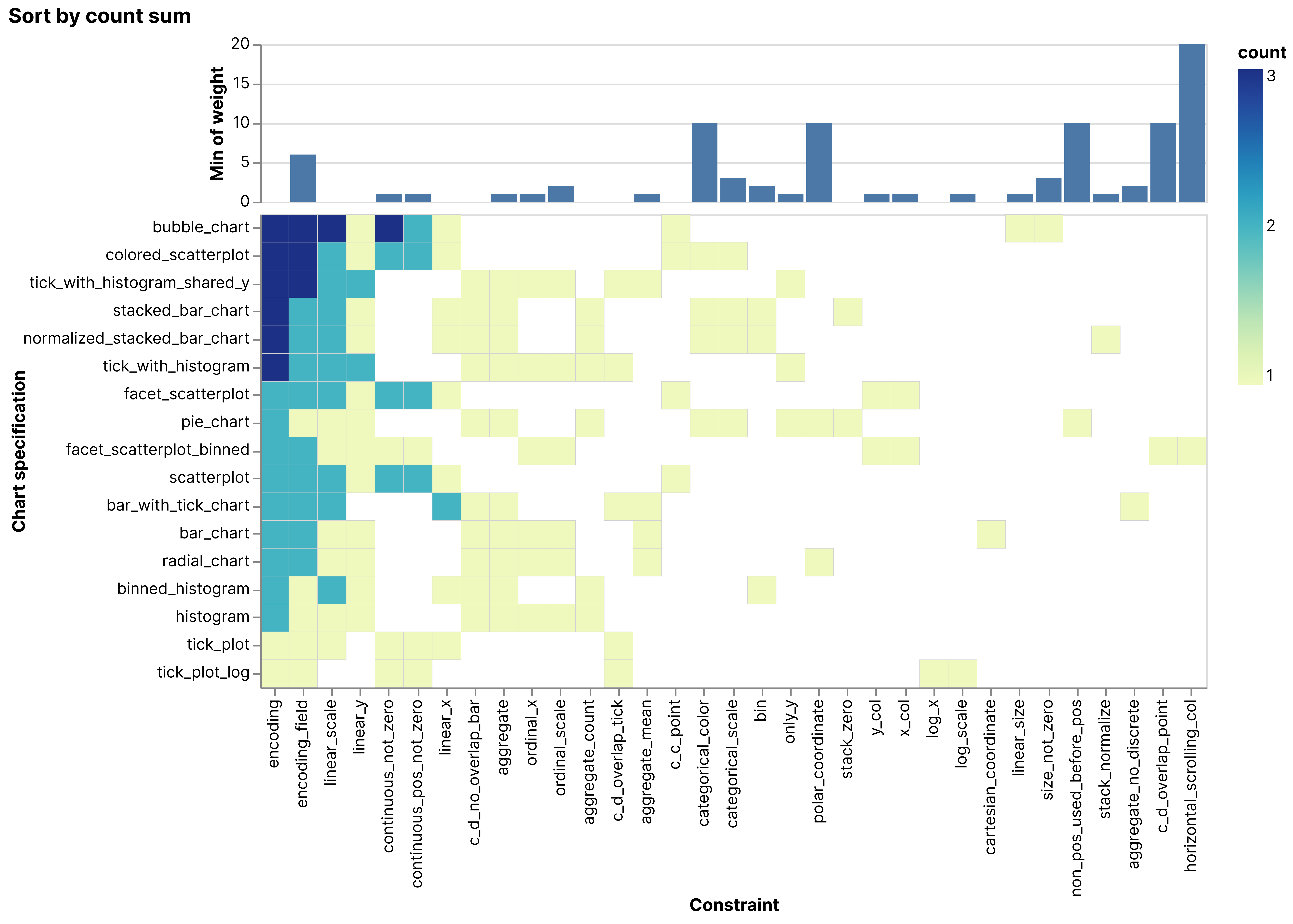}
    \vspace{-2em}
    \caption{visualizing the violation vectors with \mintinline{python}{debug.DracoDebugPlotter.create_chart}. It consists of an aligned bar chart displaying constraint weights and a heatmap illustrating constraint violation frequencies for each chart specification. }
    \label{fig:debug:heatmap}
    \vspace{-1em}
\end{figure}

\textbf{Debugging and Constraint Weight Tuning Support:} 
To revise soft constraints and tune weights, the first step is to understand the existing knowledge base by how it is reflected in the recommended results. To support interpreting and inspecting Draco's output, we provide a \textit{debugger module (\mintinline{python}{debug.DracoDebug}}) to examine which soft constraints in the knowledge base are violated for a collection of visualizations, and a \textit{plotter module (\mintinline{python}{debug.DracoDebugPlotter})} to visualize the violation vectors as shown in \autoref{fig:debug:heatmap}.
One can use these components to determine how to adjust the knowledge base to yield more optimal recommendations, whether by adding or removing constraints, or fine-tuning the weights.



%

\textbf{Web Integration:} We made Draco easily web-integrable through two main approaches: a) Deploy it as a standalone Python app interfacing via the \mintinline{python}{server} module's web API, or b) Use the Python API directly on the front-end through our WebAssembly distribution with the \href{https://www.npmjs.com/package/draco-pyodide}{\mintinline{latex}{draco-pyodide}}\footnote{\href{https://www.npmjs.com/package/draco-pyodide}{npmjs.com/package/draco-pyodide}} package.

\subsection{Comparison of Draco~1 and Draco~2}

We compare the API implementation, visualization specification format and behavior of Draco~1 and Draco~2 in a Jupyter Notebook\footnote{\href{https://dig.cmu.edu/draco2/applications/draco1_vs_draco2.html}{dig.cmu.edu/draco2/applications/draco1\_vs\_draco2.html} (\href{https://github.com/cmudig/draco2/blob/e2550b12ffd78c22d936027877334017f84c6327/docs/applications/draco1_vs_draco2.ipynb}{permalink})} through a series of examples.

Although similar features are supported by APIs of both versions, Draco~2 proves to be easier to integrate into the Python ecosystem, and it exposes additional features, such as a debugger module. 
The visualization specification format of the two versions differs vastly. While Draco~1 uses a format tied closely to Vega-Lite, Draco~2 adopts an extensible, visualization-grammar-independent format, built around the idea of using generic entities and attributes.

To compare the behavior of Draco~1 and Draco~2, we selected 100 visualization pairs from the dataset compiled by Kim and Heer \cite{kim2018assessing} for graphical perception study and let the systems rank them using their default knowledge base and constraint weights. 
Note that even slight weight changes to the knowledge base can lead to very different results, however, the two systems agreed on the ranking 86\% of the time, indicating a high behavioral similarity. In each instance of divergent rankings, Draco~2 preferred the visualization which was also deemed better by the participants of Kim and Heer's study.

\section{Demonstration of Draco as a modeling tool}

We demonstrate Draco's capabilities as an effective tool to generate recommendations, as well as to explore and adapt the knowledge base. 
We created a general guide for debugging recommendation results\footnote{\href{https://dig.cmu.edu/draco2/applications/debug_draco.html}{dig.cmu.edu/draco2/applications/debug\_draco.html}(\href{https://github.com/cmudig/draco2/blob/e2550b12ffd78c22d936027877334017f84c6327/docs/applications/debug_draco.ipynb}{permalink})}. To showcase the procedure and the debugging APIs, we created a Jupyter Notebook\footnote{\href{https://dig.cmu.edu/draco2/applications/design_space_exploration.html}{dig.cmu.edu/draco2/applications/design\_space\_exploration.html} (\href{https://github.com/cmudig/draco2/blob/e2550b12ffd78c22d936027877334017f84c6327/docs/applications/design_space_exploration.ipynb}{permalink})} that explores the visualization design space of the Seattle weather dataset\footnote{\href{https://cdn.jsdelivr.net/npm/vega-datasets@v1.29.0/data/seattle-weather.csv}{cdn.jsdelivr.net/npm/vega-datasets@v1.29.0/data/seattle-weather.csv}}. While in principle this exploration would have been possible in the original Draco~1 system, Draco~2's utilities simplify it significantly.

Draco generates visualization recommendations from incomplete specifications through the \mintinline{python}{Draco.complete_spec} method. 
The results might be unsatisfactory for several reasons. When the hard constraints are not adequate, Draco might output ill-defined results. When the hard constraints are too strong, good candidates might be filtered out without a chance to be ranked with others. Similarly, since the soft constraints and their costs determine the ranking of candidates, there needs to be a suitable degree of differentiation to distinguish between the candidates. With the following example, we demonstrate how to iteratively explore and adjust the Draco knowledge base by modifying the queries and the rules. 


\textbf{Iterating the partial specification query:} 
We start with only the raw dataset input, modeling the common first step in visual data exploration. We load the dataset and generate its schema (name of the columns, their data types, and key statistical properties) as logical facts with the \mintinline{python}{schema_from_dataframe} function. Then, for the recommendation query, we concatenate the schema with \mintinline{prolog}{entity(view,root,v0)} and \mintinline{prolog}{entity(mark,v0,m0).}, making sure that the recommendations have at least one view and use at least one mark. We generate and render the top recommendations using \mintinline{python}{complete_spec} and \mintinline{python}{AltairRenderer}. 
The top five results encode the count of records because in general using less encoding is preferred, and they represent an overview given that less information is specified by the query. Two of them encode a non-aggregated second field, \mintinline{latex}{weather}, via columnar faceting. We also observe seemingly identical charts with different costs. These are caused by the fact that there are some entities in the Draco specification such as \mintinline{python}{"task"} whose value influences the cost computed by Clingo, but does not affect the rendered Vega-Lite specification.

After an initial overview and inspiration, we explore how Draco can be used to create more targeted outputs. To constrain the desired design space, we extend the base input specification with additional facts to target \mintinline{latex}{date} temporal field and \mintinline{latex}{temp_max} numeric field in the target results. We also specify a preference for column-faceted charts, while allowing our tool to determine other details. As a result, we obtain charts faceted by the \mintinline{latex}{weather} field since it is a categorical variable with low cardinality. And the faceted scatter plot without binning is preferred to the faceted tick chart where the \mintinline{latex}{temp_max} field is binned. Draco produces visualizations from partial specification based on the fundamental visualization design guidelines expressed by our knowledge base. We can test more input variations and compare their costs, for example, by forbidding the facet and adding another encoding for the \mintinline{latex}{weather} field that could be encoded in the color channel.

\textbf{Inspecting the Knowledge Base:}
To perform a more thorough analysis of the recommendations and to validate them against the design preferences (soft constraints) defined in the knowledge base, we employ Draco's debugger module. The \mintinline{python}{DracoDebug} module generates a Pandas \mintinline{latex}{DataFrame} containing the recommendations, the violated soft constraints, and their associated weights. We use \mintinline{python}{DracoDebugPlotter} to investigate the violation vector and weights interactively. We observe that only a small subset of the defined soft constraints (18 out of 147) impact the recommended charts due to the overlap between them. 

Now, we programmatically synthesize a collection of partial specifications to explore more possibilities within the design space.  
We specify a list of marks \mintinline{python}{["point", "bar", "line", "rect"]}, fields \mintinline{python}{["weather", "temp_min", "date"]} and encoding channels \mintinline{python}{["color", "shape", "size"]} and we enumerate every combination of them in the query to obtain a variety of designs. 
We observe that Draco uses binning, faceting, stacking, and aggregations such as count and mean in an attempt to recommend meaningful visualizations. We also observe that some mark-field-channel combinations such as \mintinline{python}{("line", "date", "size")} do not yield any recommendations due to violating a hard-constraint, \mintinline{prolog}{size_without_point_text} in this case, defining that encoding data using the size channel only works when using point or text as the mark. 

The debugger output indicates that 40 of the defined 147 soft constraints influenced the recommendations from our synthesized queries, revealing a different design space to target at. 
From each specification collection, we gather insights of \textit{what combinations can or cannot be rendered, do the rendered charts seem reasonable, what are their costs and violation vectors, and do the costs reflect how they should be ranked in practice}. 
Through this process, we can confirm how well the knowledge base fits our mental model. Then, either we learn new design guidelines by verifying them with credible sources, or we have detected issues to resolve so that we can improve the knowledge base. 

\textbf{Adjusting the Knowledge Base:}
Now we demonstrate how to debug constraint logic, tune weights, and discover new rules to add. 
After analyzing the costs and violation vectors with the heatmap output from the \mintinline{python}{DracoDebugPlotter}, we may notice that there are unconventional combinations (\eg \mintinline{python}{("point", "date", "size")} and \mintinline{python}{("point", "date", "shape")}) with low costs, meaning that they are likely to be ranked higher among all recommendations with the current knowledge base. By inspecting the pattern in their violation vectors and how they differ from the others, we might detect
potential errors in the soft constraint definition or weight of constraints that can be tuned to rank them lower. For example, using the \textit{date} field for \textit{color} or \textit{size} both violate the soft constraint \mintinline{prolog}{time_not_x}, which prefers to use the field of type \textit{datetime} on the x-axis. Hence, we could increase its weight and see how that affects the results. 
We also observe non-expressive designs which share common characteristics not reflected on the violation vector heatmap. This might indicate design space that the existing knowledge base has yet to cover. For example, Draco recommended faceted heatmap designs for the \mintinline{python}{"rect"} and \mintinline{python}{"color"} combination, and the design is not discouraged by any existing soft constraints. As we decide to add such a soft constraint, we first assign it a low weight and continue to increase it to test the changes from re-runs.

In conclusion, we demonstrated how Draco can be an easy-to-use modeling tool to interact with the knowledge base and generate visualization recommendations.

\section{Future Work and Conclusion}

We believe that Draco is a timely contribution to the visualization research community.
And this implementation gets us closer to Draco's original vision of an evolving knowledge base that can be refined, extended, and tested by researchers and practitioners.
Draco is already being used as a platform for research. Zeng et al.~\cite{Zeng2023toomanycooks} used Draco to analyze the implication of different graphical perception studies.
Recent Large Language Models (LLMs) open new opportunities for visualization recommendation with natural language~\cite{dibia2023lida}. Unfortunately, the recommendations from LLMs are hard to explain, steer, and debug.
To decouple the interpretation of the natural language input and recommendation, we could use an LLM and Draco together. The LLM could generate a partial specification and then Draco could complete the specification and generate a visualization.
Another application of Draco could be as a way to embed visualizations in machine learning models. The violation of soft constraints of a particular visualization forms a vector that could be used as a feature in a machine learning model.
These are just some of the potential projects we envision for Draco: others may extend its model to fine-grained task taxonomies or interactive charts. 

In conclusion, we present a major improvement over Draco to make it more self-contained, well-documented and extensible.
We believe that these improvements make Draco a solid foundation for future research.



\acknowledgments{
We thank our labs for their feedback on this system and paper, especially Manfred Klaffenböck, Torsten Möller, Halden Lin and Ameya Patil.
}

\bibliographystyle{abbrv-doi-hyperref}

\bibliography{reference}

\begin{thebibliography}{10}

\bibitem{9552878}
\href{https://doi.org/10.1109/TVCG.2021.3114804}{Q.~Chen, F.~Sun, X.~Xu,
  Z.~Chen, J.~Wang, and N.~Cao}.
\newblock \href{https://doi.org/10.1109/TVCG.2021.3114804}{Vizlinter: A linter
  and fixer framework for data visualization}.
\newblock \href{https://doi.org/10.1109/TVCG.2021.3114804}{{\em IEEE
  Transactions on Visualization and Computer Graphics}},
  \href{https://doi.org/10.1109/TVCG.2021.3114804}{28(1):206--216},
  \href{https://doi.org/10.1109/TVCG.2021.3114804}{2022}.
  \href{https://doi.org/10.1109/TVCG.2021.3114804}
{doi: {{%
10\hspace{.1pt}\discretionary{.}{%
}{.}\hspace{.4pt}1109\discretionary{/}{%
}{/}TVCG\hspace{.1pt}\discretionary{.}{%
}{.}\hspace{.4pt}2021\hspace{.1pt}\discretionary{.}{%
}{.}\hspace{.4pt}3114804}}}


\bibitem{Demiralp2017foresight}
\href{https://doi.org/10.14778/3137765.3137813}{{\c{C}}.~Demiralp, P.~J. Haas,
  S.~Parthasarathy, and T.~Pedapati}.
\newblock \href{https://doi.org/10.14778/3137765.3137813}{Foresight:
  Recommending visual insights}.
\newblock \href{https://doi.org/10.14778/3137765.3137813}{{\em Proceedings of
  the VLDB Endowment}},
  \href{https://doi.org/10.14778/3137765.3137813}{10(12):1937–1940},
  \href{https://doi.org/10.14778/3137765.3137813}{Aug. 2017}.
  \href{https://doi.org/10.14778/3137765.3137813}
{doi: {{%
10\hspace{.1pt}\discretionary{.}{%
}{.}\hspace{.4pt}14778\discretionary{/}{%
}{/}3137765\hspace{.1pt}\discretionary{.}{%
}{.}\hspace{.4pt}3137813}}}


\bibitem{dibia2023lida}
V.~Dibia.
\newblock Lida: A tool for automatic generation of grammar-agnostic
  visualizations and infographics using large language models.
\newblock {\em arXiv preprint arXiv:2303.02927}, 2023.

\bibitem{Gebser2014clingo}
\href{http://arxiv.org/abs/1405.3694}{M.~Gebser, R.~Kaminski, B.~Kaufmann, and
  T.~Schaub}.
\newblock \href{http://arxiv.org/abs/1405.3694}{Clingo = {ASP} + control:
  Preliminary report}.
\newblock \href{http://arxiv.org/abs/1405.3694}{{\em CoRR}},
  \href{http://arxiv.org/abs/1405.3694}{abs/1405.3694},
  \href{http://arxiv.org/abs/1405.3694}{2014}.

\bibitem{Gebser2011asp}
\href{https://doi.org/10.1017/S1471068411000329}{M.~Gebser, R.~Kaminski, and
  T.~Schaub}.
\newblock \href{https://doi.org/10.1017/S1471068411000329}{Complex optimization
  in answer set programming}.
\newblock \href{https://doi.org/10.1017/S1471068411000329}{{\em Theory and
  Practice of Logic Programming}},
  \href{https://doi.org/10.1017/S1471068411000329}{11(4-5):821–839},
  \href{https://doi.org/10.1017/S1471068411000329}{2011}.
  \href{https://doi.org/10.1017/S1471068411000329}
{doi: {{%
10\hspace{.1pt}\discretionary{.}{%
}{.}\hspace{.4pt}1017\discretionary{/}{%
}{/}S1471068411000329}}}


\bibitem{Gebser2011potassco}
\href{https://doi.org/10.3233/AIC-2011-0491}{M.~Gebser, B.~Kaufmann,
  R.~Kaminski, M.~Ostrowski, T.~Schaub, and M.~Schneider}.
\newblock \href{https://doi.org/10.3233/AIC-2011-0491}{Potassco: The potsdam
  answer set solving collection}.
\newblock \href{https://doi.org/10.3233/AIC-2011-0491}{{\em AI
  Communications}},
  \href{https://doi.org/10.3233/AIC-2011-0491}{24(2):107–124},
  \href{https://doi.org/10.3233/AIC-2011-0491}{apr 2011}.
  \href{https://doi.org/10.3233/AIC-2011-0491}
{doi: {{%
10\hspace{.1pt}\discretionary{.}{%
}{.}\hspace{.4pt}3233\discretionary{/}{%
}{/}AIC\discretionary{%
}{-}{-}2011\discretionary{%
}{-}{-}0491}}}


\bibitem{hopkins2020visualint}
A.~K. Hopkins, M.~Correll, and A.~Satyanarayan.
\newblock Visualint: Sketchy in situ annotations of chart construction errors.
\newblock In {\em Computer Graphics Forum}, vol.~39, pp. 219--228. Wiley Online
  Library, 2020.

\bibitem{Hu2019vizml}
\href{https://doi.org/10.1145/3290605.3300358}{K.~Hu, M.~Bakker, S.~Li,
  T.~Kraska, and C.~Hidalgo}.
\newblock \href{https://doi.org/10.1145/3290605.3300358}{Viz{ML}: A machine
  learning approach to visualization recommendation}.
\newblock \href{https://doi.org/10.1145/3290605.3300358}{In {\em Proceedings of
  the SIGCHI Conference on Human Factors in Computing Systems}},
  \href{https://doi.org/10.1145/3290605.3300358}{CHI ’19},
  \href{https://doi.org/10.1145/3290605.3300358}{pp. 1--12}.
  \href{https://doi.org/10.1145/3290605.3300358}{Association for Computing
  Machinery}, \href{https://doi.org/10.1145/3290605.3300358}{New York, NY,
  USA}, \href{https://doi.org/10.1145/3290605.3300358}{2019}.
  \href{https://doi.org/10.1145/3290605.3300358}
{doi: {{%
10\hspace{.1pt}\discretionary{.}{%
}{.}\hspace{.4pt}1145\discretionary{/}{%
}{/}3290605\hspace{.1pt}\discretionary{.}{%
}{.}\hspace{.4pt}3300358}}}


\bibitem{Key2012vizdeck}
\href{https://doi.org/10.1145/2213836.2213931}{A.~Key, B.~Howe, D.~Perry, and
  C.~Aragon}.
\newblock \href{https://doi.org/10.1145/2213836.2213931}{Viz{D}eck:
  Self-organizing dashboards for visual analytics}.
\newblock \href{https://doi.org/10.1145/2213836.2213931}{In {\em Proceedings of
  the ACM SIGMOD International Conference on Management of Data}},
  \href{https://doi.org/10.1145/2213836.2213931}{SIGMOD '12},
  \href{https://doi.org/10.1145/2213836.2213931}{p. 681–684}.
  \href{https://doi.org/10.1145/2213836.2213931}{Association for Computing
  Machinery}, \href{https://doi.org/10.1145/2213836.2213931}{New York, NY,
  USA}, \href{https://doi.org/10.1145/2213836.2213931}{2012}.
  \href{https://doi.org/10.1145/2213836.2213931}
{doi: {{%
10\hspace{.1pt}\discretionary{.}{%
}{.}\hspace{.4pt}1145\discretionary{/}{%
}{/}2213836\hspace{.1pt}\discretionary{.}{%
}{.}\hspace{.4pt}2213931}}}


\bibitem{kim2018assessing}
\href{https://doi.org/https://doi.org/10.1111/cgf.13409}{Y.~Kim and J.~Heer}.
\newblock \href{https://doi.org/https://doi.org/10.1111/cgf.13409}{Assessing
  effects of task and data distribution on the effectiveness of visual
  encodings}.
\newblock \href{https://doi.org/https://doi.org/10.1111/cgf.13409}{{\em
  Computer Graphics Forum}},
  \href{https://doi.org/https://doi.org/10.1111/cgf.13409}{37(3):157--167},
  \href{https://doi.org/https://doi.org/10.1111/cgf.13409}{2018}.
  \href{https://doi.org/10.1111/cgf.13409}
{doi: {{%
10\hspace{.1pt}\discretionary{.}{%
}{.}\hspace{.4pt}1111\discretionary{/}{%
}{/}cgf\hspace{.1pt}\discretionary{.}{%
}{.}\hspace{.4pt}13409}}}


\bibitem{Li2022kg4vis}
\href{https://doi.org/10.1109/TVCG.2021.3114863}{H.~Li, Y.~Wang, S.~Zhang,
  Y.~Song, and H.~Qu}.
\newblock \href{https://doi.org/10.1109/TVCG.2021.3114863}{{KG}4{V}is: A
  knowledge graph-based approach for visualization recommendation}.
\newblock \href{https://doi.org/10.1109/TVCG.2021.3114863}{{\em IEEE
  Transactions on Visualization and Computer Graphics}},
  \href{https://doi.org/10.1109/TVCG.2021.3114863}{28(1):195--205},
  \href{https://doi.org/10.1109/TVCG.2021.3114863}{2022}.
  \href{https://doi.org/10.1109/TVCG.2021.3114863}
{doi: {{%
10\hspace{.1pt}\discretionary{.}{%
}{.}\hspace{.4pt}1109\discretionary{/}{%
}{/}TVCG\hspace{.1pt}\discretionary{.}{%
}{.}\hspace{.4pt}2021\hspace{.1pt}\discretionary{.}{%
}{.}\hspace{.4pt}3114863}}}


\bibitem{Luo2018deepeye}
\href{https://doi.org/10.1109/ICDE.2018.00019}{Y.~{Luo}, X.~{Qin}, N.~{Tang},
  and G.~{Li}}.
\newblock \href{https://doi.org/10.1109/ICDE.2018.00019}{Deep{E}ye: Towards
  automatic data visualization}.
\newblock \href{https://doi.org/10.1109/ICDE.2018.00019}{In {\em 2018 IEEE 34th
  International Conference on Data Engineering (ICDE)}},
  \href{https://doi.org/10.1109/ICDE.2018.00019}{pp. 101--112},
  \href{https://doi.org/10.1109/ICDE.2018.00019}{April 2018}.
  \href{https://doi.org/10.1109/ICDE.2018.00019}
{doi: {{%
10\hspace{.1pt}\discretionary{.}{%
}{.}\hspace{.4pt}1109\discretionary{/}{%
}{/}ICDE\hspace{.1pt}\discretionary{.}{%
}{.}\hspace{.4pt}2018\hspace{.1pt}\discretionary{.}{%
}{.}\hspace{.4pt}00019}}}


\bibitem{Mackinlay1986automating}
\href{https://doi.org/10.1145/22949.22950}{J.~Mackinlay}.
\newblock \href{https://doi.org/10.1145/22949.22950}{Automating the design of
  graphical presentations of relational information}.
\newblock \href{https://doi.org/10.1145/22949.22950}{{\em ACM Transactions on
  Graphics}}, \href{https://doi.org/10.1145/22949.22950}{5(2):110–141},
  \href{https://doi.org/10.1145/22949.22950}{Apr. 1986}.
  \href{https://doi.org/10.1145/22949.22950}
{doi: {{%
10\hspace{.1pt}\discretionary{.}{%
}{.}\hspace{.4pt}1145\discretionary{/}{%
}{/}22949\hspace{.1pt}\discretionary{.}{%
}{.}\hspace{.4pt}22950}}}


\bibitem{Mackinlay2007showme}
\href{https://doi.org/10.1109/TVCG.2007.70594}{J.~Mackinlay, P.~Hanrahan, and
  C.~Stolte}.
\newblock \href{https://doi.org/10.1109/TVCG.2007.70594}{Show {M}e: Automatic
  presentation for visual analysis}.
\newblock \href{https://doi.org/10.1109/TVCG.2007.70594}{{\em IEEE Transactions
  on Visualization and Computer Graphics}},
  \href{https://doi.org/10.1109/TVCG.2007.70594}{13(6):1137--1144},
  \href{https://doi.org/10.1109/TVCG.2007.70594}{Nov 2007}.
  \href{https://doi.org/10.1109/TVCG.2007.70594}
{doi: {{%
10\hspace{.1pt}\discretionary{.}{%
}{.}\hspace{.4pt}1109\discretionary{/}{%
}{/}TVCG\hspace{.1pt}\discretionary{.}{%
}{.}\hspace{.4pt}2007\hspace{.1pt}\discretionary{.}{%
}{.}\hspace{.4pt}70594}}}


\bibitem{McNutt2018LintingFV}
A.~M. McNutt and G.~L. Kindlmann.
\newblock Linting for visualization: Towards a practical automated
  visualization guidance system.
\newblock 2018.

\bibitem{Moritz2018formalizing}
\href{https://doi.org/10.1109/TVCG.2018.2865240}{D.~Moritz, C.~Wang, G.~L.
  Nelson, H.~Lin, A.~M. Smith, B.~Howe, and J.~Heer}.
\newblock \href{https://doi.org/10.1109/TVCG.2018.2865240}{Formalizing
  visualization design knowledge as constraints: Actionable and extensible
  models in draco}.
\newblock \href{https://doi.org/10.1109/TVCG.2018.2865240}{{\em IEEE
  Transactions on Visualization and Computer Graphics}},
  \href{https://doi.org/10.1109/TVCG.2018.2865240}{25(1):438--448},
  \href{https://doi.org/10.1109/TVCG.2018.2865240}{Jan 2019}.
  \href{https://doi.org/10.1109/TVCG.2018.2865240}
{doi: {{%
10\hspace{.1pt}\discretionary{.}{%
}{.}\hspace{.4pt}1109\discretionary{/}{%
}{/}TVCG\hspace{.1pt}\discretionary{.}{%
}{.}\hspace{.4pt}2018\hspace{.1pt}\discretionary{.}{%
}{.}\hspace{.4pt}2865240}}}


\bibitem{Satyanarayan2017vegalite}
\href{https://doi.org/10.1109/TVCG.2016.2599030}{A.~{Satyanarayan},
  D.~{Moritz}, K.~{Wongsuphasawat}, and J.~{Heer}}.
\newblock \href{https://doi.org/10.1109/TVCG.2016.2599030}{Vega-lite: A grammar
  of interactive graphics}.
\newblock \href{https://doi.org/10.1109/TVCG.2016.2599030}{{\em IEEE
  Transactions on Visualization and Computer Graphics}},
  \href{https://doi.org/10.1109/TVCG.2016.2599030}{23(1):341--350},
  \href{https://doi.org/10.1109/TVCG.2016.2599030}{Jan 2017}.
  \href{https://doi.org/10.1109/TVCG.2016.2599030}
{doi: {{%
10\hspace{.1pt}\discretionary{.}{%
}{.}\hspace{.4pt}1109\discretionary{/}{%
}{/}TVCG\hspace{.1pt}\discretionary{.}{%
}{.}\hspace{.4pt}2016\hspace{.1pt}\discretionary{.}{%
}{.}\hspace{.4pt}2599030}}}


\bibitem{Siddiqui2017fast}
\href{http://cidrdb.org/cidr2017/papers/p43-siddiqui-cidr17.pdf}{T.~Siddiqui,
  J.~Lee, A.~Kim, E.~Xue, X.~Yu, S.~Zou, L.~Guo, C.~Liu, C.~Wang,
  K.~Karahalios, and A.~G. Parameswaran}.
\newblock
  \href{http://cidrdb.org/cidr2017/papers/p43-siddiqui-cidr17.pdf}{Fast-forwarding
  to desired visualizations with zenvisage}.
\newblock \href{http://cidrdb.org/cidr2017/papers/p43-siddiqui-cidr17.pdf}{In
  {\em 8th Biennial Conference on Innovative Data Systems Research, {CIDR}
  2017, Chaminade, CA, USA, January 8-11, 2017, Online Proceedings}}.
  \href{http://cidrdb.org/cidr2017/papers/p43-siddiqui-cidr17.pdf}{www.cidrdb.org},
  \href{http://cidrdb.org/cidr2017/papers/p43-siddiqui-cidr17.pdf}{2017}.

\bibitem{Vartak2015seedb}
\href{https://doi.org/10.14778/2831360.2831371}{M.~Vartak, S.~Rahman,
  S.~Madden, A.~Parameswaran, and N.~Polyzotis}.
\newblock \href{https://doi.org/10.14778/2831360.2831371}{See{DB}: Efficient
  data-driven visualization recommendations to support visual analytics}.
\newblock \href{https://doi.org/10.14778/2831360.2831371}{{\em Proceedings of
  the VLDB Endowment}},
  \href{https://doi.org/10.14778/2831360.2831371}{8(13):2182–2193},
  \href{https://doi.org/10.14778/2831360.2831371}{Sept. 2015}.
  \href{https://doi.org/10.14778/2831360.2831371}
{doi: {{%
10\hspace{.1pt}\discretionary{.}{%
}{.}\hspace{.4pt}14778\discretionary{/}{%
}{/}2831360\hspace{.1pt}\discretionary{.}{%
}{.}\hspace{.4pt}2831371}}}


\bibitem{wilkinson2005gog}
L.~Wilkinson.
\newblock {\em The Grammar of Graphics (Statistics and Computing)}.
\newblock Springer-Verlag, Berlin, Heidelberg, 2005.

\bibitem{Wongsuphasawat2016towards}
\href{https://doi.org/10.1145/2939502.2939506}{K.~Wongsuphasawat, D.~Moritz,
  A.~Anand, J.~Mackinlay, B.~Howe, and J.~Heer}.
\newblock \href{https://doi.org/10.1145/2939502.2939506}{Towards a
  general-purpose query language for visualization recommendation}.
\newblock \href{https://doi.org/10.1145/2939502.2939506}{In {\em Proceedings of
  the Workshop on Human-In-the-Loop Data Analytics}},
  \href{https://doi.org/10.1145/2939502.2939506}{HILDA '16},
  \href{https://doi.org/10.1145/2939502.2939506}{pp. 4:1--4:6}.
  \href{https://doi.org/10.1145/2939502.2939506}{ACM},
  \href{https://doi.org/10.1145/2939502.2939506}{New York, NY, USA},
  \href{https://doi.org/10.1145/2939502.2939506}{2016}.
  \href{https://doi.org/10.1145/2939502.2939506}
{doi: {{%
10\hspace{.1pt}\discretionary{.}{%
}{.}\hspace{.4pt}1145\discretionary{/}{%
}{/}2939502\hspace{.1pt}\discretionary{.}{%
}{.}\hspace{.4pt}2939506}}}


\bibitem{Wongsuphasawat2015voyager}
\href{https://doi.org/10.1109/TVCG.2015.2467191}{K.~Wongsuphasawat, D.~Moritz,
  A.~Anand, J.~Mackinlay, B.~Howe, and J.~Heer}.
\newblock \href{https://doi.org/10.1109/TVCG.2015.2467191}{Voyager: Exploratory
  analysis via faceted browsing of visualization recommendations}.
\newblock \href{https://doi.org/10.1109/TVCG.2015.2467191}{vol.~22},
  \href{https://doi.org/10.1109/TVCG.2015.2467191}{pp. 649--658},
  \href{https://doi.org/10.1109/TVCG.2015.2467191}{Jan 2016}.
  \href{https://doi.org/10.1109/TVCG.2015.2467191}
{doi: {{%
10\hspace{.1pt}\discretionary{.}{%
}{.}\hspace{.4pt}1109\discretionary{/}{%
}{/}TVCG\hspace{.1pt}\discretionary{.}{%
}{.}\hspace{.4pt}2015\hspace{.1pt}\discretionary{.}{%
}{.}\hspace{.4pt}2467191}}}


\bibitem{Wongsuphasawat2017voyager2}
\href{https://doi.org/10.1145/3025453.3025768}{K.~Wongsuphasawat, Z.~Qu,
  D.~Moritz, R.~Chang, F.~Ouk, A.~Anand, J.~Mackinlay, B.~Howe, and J.~Heer}.
\newblock \href{https://doi.org/10.1145/3025453.3025768}{Voyager 2: Augmenting
  visual analysis with partial view specifications}.
\newblock \href{https://doi.org/10.1145/3025453.3025768}{In {\em Proceedings of
  the SIGCHI Conference on Human Factors in Computing Systems}},
  \href{https://doi.org/10.1145/3025453.3025768}{CHI ’17},
  \href{https://doi.org/10.1145/3025453.3025768}{p. 2648–2659}.
  \href{https://doi.org/10.1145/3025453.3025768}{Association for Computing
  Machinery}, \href{https://doi.org/10.1145/3025453.3025768}{New York, NY,
  USA}, \href{https://doi.org/10.1145/3025453.3025768}{2017}.
  \href{https://doi.org/10.1145/3025453.3025768}
{doi: {{%
10\hspace{.1pt}\discretionary{.}{%
}{.}\hspace{.4pt}1145\discretionary{/}{%
}{/}3025453\hspace{.1pt}\discretionary{.}{%
}{.}\hspace{.4pt}3025768}}}


\bibitem{Zeng2023review}
\href{https://doi.org/10.1145/3544548.3581349}{Z.~Zeng and L.~Battle}.
\newblock \href{https://doi.org/10.1145/3544548.3581349}{A review and collation
  of graphical perception knowledge for visualization recommendation}.
\newblock \href{https://doi.org/10.1145/3544548.3581349}{In {\em Proceedings of
  the SIGCHI Conference on Human Factors in Computing Systems}},
  \href{https://doi.org/10.1145/3544548.3581349}{CHI '23}.
  \href{https://doi.org/10.1145/3544548.3581349}{ACM},
  \href{https://doi.org/10.1145/3544548.3581349}{New York, NY, USA},
  \href{https://doi.org/10.1145/3544548.3581349}{2023}.
  \href{https://doi.org/10.1145/3544548.3581349}
{doi: {{%
10\hspace{.1pt}\discretionary{.}{%
}{.}\hspace{.4pt}1145\discretionary{/}{%
}{/}3544548\hspace{.1pt}\discretionary{.}{%
}{.}\hspace{.4pt}3581349}}}


\bibitem{Zeng2021evaluation}
\href{https://doi.org/10.1109/TVCG.2021.3114814}{Z.~Zeng, P.~Moh, F.~Du,
  J.~Hoffswell, T.~Y. Lee, S.~Malik, E.~Koh, and L.~Battle}.
\newblock \href{https://doi.org/10.1109/TVCG.2021.3114814}{An
  evaluation-focused framework for visualization recommendation algorithms}.
\newblock \href{https://doi.org/10.1109/TVCG.2021.3114814}{{\em IEEE
  Transactions on Visualization and Computer Graphics}},
  \href{https://doi.org/10.1109/TVCG.2021.3114814}{28(1):346--356},
  \href{https://doi.org/10.1109/TVCG.2021.3114814}{2022}.
  \href{https://doi.org/10.1109/TVCG.2021.3114814}
{doi: {{%
10\hspace{.1pt}\discretionary{.}{%
}{.}\hspace{.4pt}1109\discretionary{/}{%
}{/}TVCG\hspace{.1pt}\discretionary{.}{%
}{.}\hspace{.4pt}2021\hspace{.1pt}\discretionary{.}{%
}{.}\hspace{.4pt}3114814}}}


\bibitem{Zeng2023toomanycooks}
Z.~Zeng, J.~Yang, D.~Moritz, J.~Heer, and L.~Battle.
\newblock Too many cooks: Exploring how graphical perception studies influence
  visualization recommendations in draco.
\newblock {\em IEEE Transactions on Visualization and Computer Graphics}, 2023.

\bibitem{Zhu2020survey}
\href{https://doi.org/10.1016/j.visinf.2020.07.002}{S.~Zhu, G.~Sun, Q.~Jiang,
  M.~Zha, and R.~Liang}.
\newblock \href{https://doi.org/10.1016/j.visinf.2020.07.002}{A survey on
  automatic infographics and visualization recommendations}.
\newblock \href{https://doi.org/10.1016/j.visinf.2020.07.002}{{\em Visual
  Informatics}},
  \href{https://doi.org/10.1016/j.visinf.2020.07.002}{4(3):24--40},
  \href{https://doi.org/10.1016/j.visinf.2020.07.002}{2020}.
  \href{https://doi.org/10.1016/j.visinf.2020.07.002}
{doi: {{%
10\hspace{.1pt}\discretionary{.}{%
}{.}\hspace{.4pt}1016\discretionary{/}{%
}{/}j\hspace{.1pt}\discretionary{.}{%
}{.}\hspace{.4pt}visinf\hspace{.1pt}\discretionary{.}{%
}{.}\hspace{.4pt}2020\hspace{.1pt}\discretionary{.}{%
}{.}\hspace{.4pt}07\hspace{.1pt}\discretionary{.}{%
}{.}\hspace{.4pt}002}}}


\end{thebibliography}
\end{document}